\documentclass[fleqn,usenatbib]{mnras}

\usepackage{newtxtext,newtxmath}

\usepackage[T1]{fontenc}
\usepackage{ae,aecompl}

\usepackage{graphicx}	
\usepackage{amsmath}	
\usepackage{amssymb}	
\usepackage{array}	

\newcolumntype{A}{>{\centering\arraybackslash} m{0.18\columnwidth} }
\newcolumntype{B}{ m{0.25\columnwidth} }

\title[Origin of Triton's cratering asymmetry]{The origin of the cratering asymmetry on Triton}

\author[J. Mah and R. Brasser]{
Jingyi Mah and Ramon Brasser\thanks{E-mail:brasser\_astro@yahoo.com}
\\
Earth Life Science Institute, Meguro-ku, Tokyo, Japan 152-8550
}

\date{Accepted 2019 March 15. Received 2019 March 13; in original form 2019 January 10}

\pubyear{2019}

\begin{document}
\label{firstpage}
\pagerange{\pageref{firstpage}--\pageref{lastpage}}
\maketitle

\begin{abstract}
The source of impactors capable of producing the cratering asymmetry on Triton is open to debate. 
While hyperbolic comets are the likely source of impactors, the theoretically predicted crater distribution due to impacts from comets is incompatible with the observed crater distribution on Triton.
In this study, we adopted a dynamical approach to investigate the crater distribution produced by three sources of impactors: Neptune-centric impactors on prograde, low-inclination orbits, Neptune-centric impactors on high-inclination orbits such as its irregular satellites, and heliocentric impactors with hyperbolic orbits.
We employ \textit{N}-body simulations which take into account the dynamics of the impacts.
We found that none of the impactor sources alone are able to reconcile with the crater distribution on Triton.
The crater distribution produced by a low-inclination Neptune-centric source is narrowly concentrated on Triton's leading hemisphere whereas the heliocentric impactors produce too many craters on the trailing hemisphere compared to the observed crater distribution.
All three impactor populations produce craters on the trailing hemisphere of Triton.
We derived the impact distribution of a mixed population of impactors using a Monte Carlo approach and found that if the population consists of mainly heliocentric impactors (70\%), the resulting impact distribution fits the observed crater distribution on Triton's leading hemisphere well.
\end{abstract}




\section{Introduction}
Being the only large satellite among those of the giant planets to orbit in a retrograde manner, Neptune's tidally-locked satellite, Triton, is undoubtedly unique.
The origins of this icy satellite lie in a capture process, most likely via an exchange capture due to the dissociation of a binary \citep{am2006}, shortly after Neptune formed followed by tidal evolution which circularised its primordial orbit to the one we observed today (e.g. \citealt{nogueiraetal2011}).

The study of this distant world was made possible in 1989 when \textit{Voyager 2} returned images of $\sim$35\% of Triton's surface. 
Attempts to map the crater density distribution on Triton has been made, by \cite{strometal1990} and then updated by \cite{sz2007} (hereafter SZ07). 
The earlier crater counting results suggest an asymmetry between the leading hemisphere (the side facing the direction of motion) and the trailing hemisphere of Triton.
This phenomenon is itself not a unique occurrence in the solar system, for crater asymmetry is a common trait of satellites which are tidally-locked (\citealt{sw1982,hn1984,zahnleetal2001}).
For Triton, as it was confirmed by SZ07 with higher quality images, the leading-trailing asymmetry of its crater distribution was not only real but more importantly, extreme. 
All the craters on Triton are found only on the leading hemisphere and they are distributed within 85$^{\circ}$ from the apex (direction of motion of Triton).
The trailing hemisphere is essentially devoid of craters.
This is true at least for the area of Triton imaged by \textit{Voyager}.
In addition, these studies also show that craters are sparsely distributed on this icy satellite in comparison to the other satellites of the giant planets, implying that Triton is of young geological age.
Triton's surface age is estimated to be at most $\sim50$ Myr for the terrain with craters (its leading hemisphere) and $\sim6$ Myr for the craterless terrain (its trailing hemisphere) (SZ07).

While it is generally believed that heliocentric objects, such as Kuiper Belt Objects (KBOs), with highly elliptical or hyperbolic orbits are the main source of impactors for bodies in the outer Solar System, the most updated crater density distribution on Triton suggests otherwise.
The crater distribution when fitted with the predicted impact distribution produced by a low-inclination and prograde planetocentric impactor source turned out to be a good match (SZ07).
This conclusion is in contrast with the results from earlier studies on Triton's craters which favoured external, i.e., heliocentric sources as the cratering agent.
Using the earlier crater statistics, \cite{strometal1990} deduced that the Triton impactors are of cometary origin based on the similarity of Triton's crater distribution with that of Miranda. 
\cite{sm2000} arrived at the same conclusion by estimating the cratering fluxes of KBOs.
We have grounds to believe that the results of the more recent work by SZ07 is more precise due to improved image quality, however, the main drawback of the conclusion that planetocentric impactors with prograde and low-inclination orbits better explain the crater distribution on Triton is that there is no evidence for the existence of such a population around Neptune.
Moreover, it is difficult to explain the existence of such a population in the past without invoking the destruction of some prograde satellite of Neptune.

A consensus is therefore lacking for the source of impactors which resulted in the cratering asymmetry on Triton.
In this paper, we revisit this problem to determine if the observed crater distribution on Triton could be explained by taking into account impact dynamics of three distinct types of impactor source populations with Triton, namely prograde, low-inclination Neptune-centric impactors, high-inclination Neptune-centric impactors, and heliocentric impactors with hyperbolic orbits.
We also study the crater distribution produced by a mixed population of two and three types of impactors.

\section{Numerical methods}
\label{sec:methods}
To ascertain the source of impactors that strike Triton and produced its extreme cratering asymmetry, we adopted a two-step approach for our numerical investigations.
The first part is aimed at obtaining the distribution of impacts as a function of the apex angle $\theta$ on Triton's surface due to various impactor sources using \textit{N}-body simulations.
The second part utilises the results from the first part to derive the impact distribution produced by a mixed impactor source with various combinations of mixing ratios via a Monte Carlo method.

\subsection{\textit{N}-body simulations}
We carried out \textit{N}-body simulations to study the dynamics of the Neptune-Triton system with three different types of potential impactor populations: (i) Neptune-centric impactors with prograde and low-inclination orbits, (ii) Neptune-centric impactors with orbits akin to the irregular satellites, and (iii) heliocentric impactors with hyperbolic orbits.
The simulations employed the regularized mixed-variable symplectic integration scheme \citep{ld1994} capable of handling close encounters.
In our simulations, the impactors are represented by massless test particles, introduced at the beginning of the simulations.
When there is a collision between a test particle and a massive body, the integrator records the vectors of both the test particle and the massive body at the time of collision before removing the test particle from the simulation.
The set up and initial conditions of the simulations are described in the following.

\textit{Prograde, low-inclination Neptune-centric impactors:} We first consider the scenario where the impactor source is a flat disc of debris (orbital inclination $I = 0$) orbiting Neptune in the prograde direction.
The periapses $q$ and apoapses $Q$ of the test particles were selected at random and constrained to fulfill the following condition such that their orbits are guaranteed to cross that of Triton:
\begin{equation}
7\ R_N \leq q \leq 14\ R_N,\ 14\ R_N \leq Q \leq 21\ R_N,
\end{equation}
where $R_N$ is Neptune radius.
The remaining orbital elements, i.e., longitude of ascending node $\Omega$, argument of periapsis $\omega$, and mean anomaly $\mathcal{M}$, of each test particle were assigned values randomly from 0$^{\circ}$ to 360$^{\circ}$.
Triton's semi-major axis, eccentricity, and inclination were fixed at $a_T = 14.41\ R_N$, $e_T = 0.0$, and $I_T = 156.83^{\circ}$ for all of the simulations carried out in this work and its phase was randomised to avoid selection biases.
We ran 100 simulations each with 100 test particles for 2.5 $\times$ 10$^5$ yr with a timestep of 0.073 d (equivalent to dividing the orbit of Triton into 80 steps).
The condition for the removal of the test particles are either (a) collision with Triton or Neptune, (b) their semi-major axes are less than 4.83 $R_N$, or (c) their semi-major axes are greater than 3625 $R_N$.
For these 100 sets of simulations, we omitted the effects from the Sun and set Neptune as the central body around which Triton and the test particles revolve.
We excluded the other Neptunian satellites in our simulations.

\textit{High-inclination Neptune-centric impactors:} We also consider the scenario where the impactors have orbits similar to the irregular satellites of Neptune.
For this scenario only, the Sun is included as a planet orbiting Neptune in the integrations.
It serves as an external perturber to deliver the high-inclination test particles (ranging from $64^{\circ} \leq I \leq 110^{\circ}$ in our simulations) to Triton via the Kozai resonance (\citealt{carrubaetal2002,nesvornyetal2003}).
The values of semi-major axes assigned to the test particles range from 0.02 au $\leq a \leq$ 0.20 au.
Both the value ranges for $I$ and $a$ were referenced from the results of the study by \cite{nesvornyetal2007} on the orbital elements of irregular satellites captured by Neptune.
We set the range of values for the periapses of the test particles to be $2\ R_N \leq q \leq 15\ R_N$.
With the values of $a$, $e$, and $I$ as inputs, the final maximum and minimum eccentricity and inclination (as well as the periapse) of a particular orbit can be solved for analytically using the method described in \cite{kn2007}.
We made use of their result to further restrict the final value of $q_{min}$ of the test particles over one Kozai cycle to be larger than 2 $R_N$ to avoid collisions with Neptune.
Once again, the three angles $\Omega$, $\omega$, and $\mathcal{M}$ were randomly selected from 0$^{\circ}$ to 360$^{\circ}$.
We carried out 90 sets of simulations with 250 test particles in each simulation.
We set the integration time and timestep to be the same as the simulations for the prograde impactors described in the preceding paragraph.
The test particles were removed if they fulfill the removal condition $a < 1\ R_N$ and $a > 211\,471\ R_N$ (or equivalently 35 au). 

\textit{Heliocentric impactors with hyperbolic orbits:} Finally we consider the scenario where the impactors have hyperbolic orbits, such as objects from the Kuiper Belt.
We obtained the encounter statistics (distribution of $q$, velocity distribution, and number of encounters) of planetesimals with Neptune from the Nice model simulations of \cite{nogueiraetal2011}.
We then selected planetesimals with orbits that have $q \leq 17 \ R_N$, chose at random $a$ and $I$ from their distributions corresponding to $q \leq 17 \ R_N$ and assigned these values to the test particles in our simulations.
We chose such initial conditions despite them originating from the past because it is known that the encounter conditions for the planets at present are similar to that during the migrating phase \citep{vokrouhlickyetal2008}.
Similar to the set up for the previous two sets of simulations, $\Omega$ and $\omega$ were chosen randomly.
The initial mean anomaly $\mathcal{M}$ was chosen to correspond to the impactor being at 1 Hill radius from Neptune on the incoming leg of the hyperbolic orbit.
As most of the test particles will continue on a path away from the Neptune-Triton system after passing through periapsis and not collide with Triton due to its small size, we increased the total number of simulations as well as the number of test particles in each simulation to obtain more impacts on Triton.
500 simulations each with 25\,000 test particles were carried out with the integration time set to 50 yr and the timestep to 0.018 d, smaller than the value set for the previous sets of simulations.
The reason for the smaller timestep is is that we want to resolve high velocity impacts.
The conditions for the removal of test particles are the same as the simulations for the high-inclination Neptune-centric impactors.

\subsection{Monte Carlo method}
Building upon the output of the first part, we generated random impacts on Triton using a combination of impactors.
We considered mixed impactor populations made up of a combination of two or three types of impactors.
As the crater density data in SZ07 is only available for $\theta = 10^{\circ}$ to 90$^{\circ}$ (c.f. their Figure 9), we restrict our attention to the impacts that occur within the same range of impact angles.
For each combination, we generated a total of 10$^3$ impacts by randomly selecting the impact angles from the impact distribution obtained from the \textit{N}-body simulations and then match the cumulative impact distribution as a function of $\theta$ with the observed crater density reported in SZ07 for craters larger than 5 km in diameter.
The aim of this part is to find out the combination of impactor types and their respective mixing ratios that would fit the observed crater distribution on Triton best.

\section{Results and discussion}
\label{sec:results_discussion}
In this section we present the results of our numerical investigations, first for the three impactor populations studied using \textit{N}-body simulations, then followed by the results for the mixed impactor populations obtained via the Monte Carlo method.

\subsection{Homogeneous impactor populations}
As the outputs (position and velocity vectors) of the simulations are in a Neptune-centric frame of reference, we have to convert them into a Triton-centric frame of reference where Triton's direction of motion is along the $+y$-axis (coinciding with the apex where $\theta = 0^{\circ}$) with Neptune located at the $+x$ direction.
This is achieved by applying a series of rotation matrices to the output vectors \citep{im2010}.

\subsubsection{Impact probabilities}
Table \ref{tab:collprob} summarises the number of collisions of each type of impactor population on Triton.
Over a course of $2.5 \times 10^5$ yr, more than 70\% of the population of disc impactors hit Triton.
The collision probability is the highest for this population because Triton encounters the swarm of impactors twice each time it completes an orbit around Neptune.
On the other hand, the collision probabilities of the other impactor populations are much lower ($<10\%$).
This is because (i) they are located further away from the Neptune-Triton system and thus spend little time near Triton, and (ii) they are at high inclination.

\begin{table}
	\centering
    \caption{Total number of test particles generated in the simulations and the resulting number of impacts on Triton for each impactor population}
    \label{tab:collprob}
    \begin{tabular}{BAAA}
    \hline
     & Disc & Irregular & Hyperbolic \\ 
    \hline
    Total no. of test particles & 10\,000 & 22\,500 & $10^7$  \\
    No. of impacts & 7\,301 & 1\,339 & 1\,557 \\
    \hline
    \end{tabular}
\end{table}

\subsubsection{Prograde, low-inclination Neptune-centric impactors}
\label{sec:disc}
\begin{figure}
	\includegraphics[width=\columnwidth]{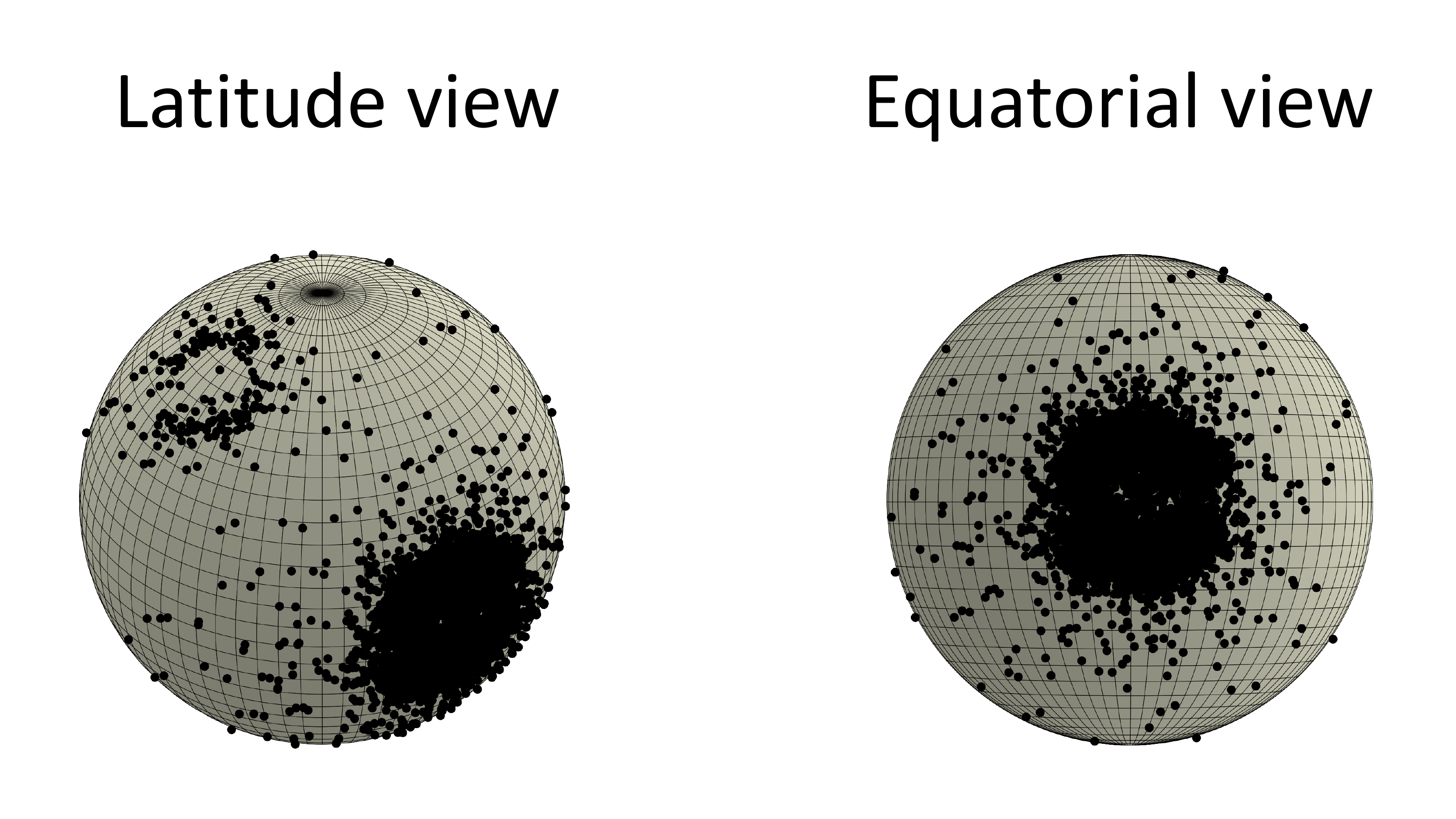}
    \caption{Distribution of impacts on the surface of Triton produced by prograde, low-inclination Neptune-centric impactors viewed from an angle (left panel) and from the equator facing Triton's leading hemisphere (right panel). The impacts on Triton's trailing hemisphere appear as an illusory ring structure near to the pole in the latitude view.}
    \label{fig:disc_surface}
\end{figure}

Figure \ref{fig:disc_surface} shows the distribution of impacts on the surface of Triton made by Neptune-centric impactors on a disc from different viewing angles.
The majority of the impacts are concentrated at a narrow circular region on the leading hemisphere surrounding the apex.
Impacts on the rest of the surface of the leading hemisphere as well as the trailing hemisphere are few and far between.

\begin{figure}
	\includegraphics[width=\columnwidth]{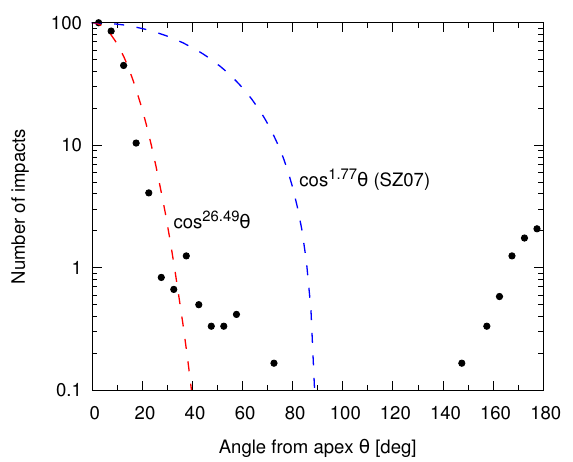}
    \caption{Projection of the impact distribution due to prograde, low-inclination Neptune-centric impactors onto the $xy$-plane. Impacts are measured with respect to the apex ($\theta = 0^{\circ}$) which coincides with Triton's direction of motion. The red dashed line is the best-fit to our numerical results. The blue dashed line is the observed crater distribution on Triton.}
    \label{fig:disc_xy}
\end{figure}

Figure \ref{fig:disc_xy} is a complementary plot of the distribution of impacts projected onto the $xy$-plane where the apex is along the $+y$-axis.
We binned the impacts into 5$^{\circ}$-wide bins and scaled the number of impacts such that the bin with the maximum number of impacts has an impact count of 100.
This facilitates the comparison of our results with the literature because the maximum crater density for $D > 5$ km is $\sim$100 (SZ07).
The population of prograde, low-inclination impactors result in a very narrow impact distribution with the majority of the impactors striking Triton within 20$^{\circ}$ from its apex.
There are some residuals at impact angles $\theta \geq \pm 160^{\circ}$ indicating that the impactors also produced impacts (although to a much smaller extent) on the trailing hemisphere of Triton which are concentrated near the antapex ($\theta = 180^{\circ}$).

An updated map of craters on Triton by SZ07 showed that the distribution of craters on Triton is best described by cosine function with an exponent of 1.77 (blue dashed line in Figure \ref{fig:disc_xy}), most likely produced by a prograde Neptune-centric source.
The results of our simulations, which account for the dynamics of the impact events, suggest that prograde, low-inclination Neptune-centric impactors (at least by themselves) could not be the sole source of craters on Triton as they result in a crater distribution that is much narrower than the actual crater distribution on Triton.

\subsubsection{High-inclination Neptune-centric impactors}
\begin{figure}
	\includegraphics[width=\columnwidth]{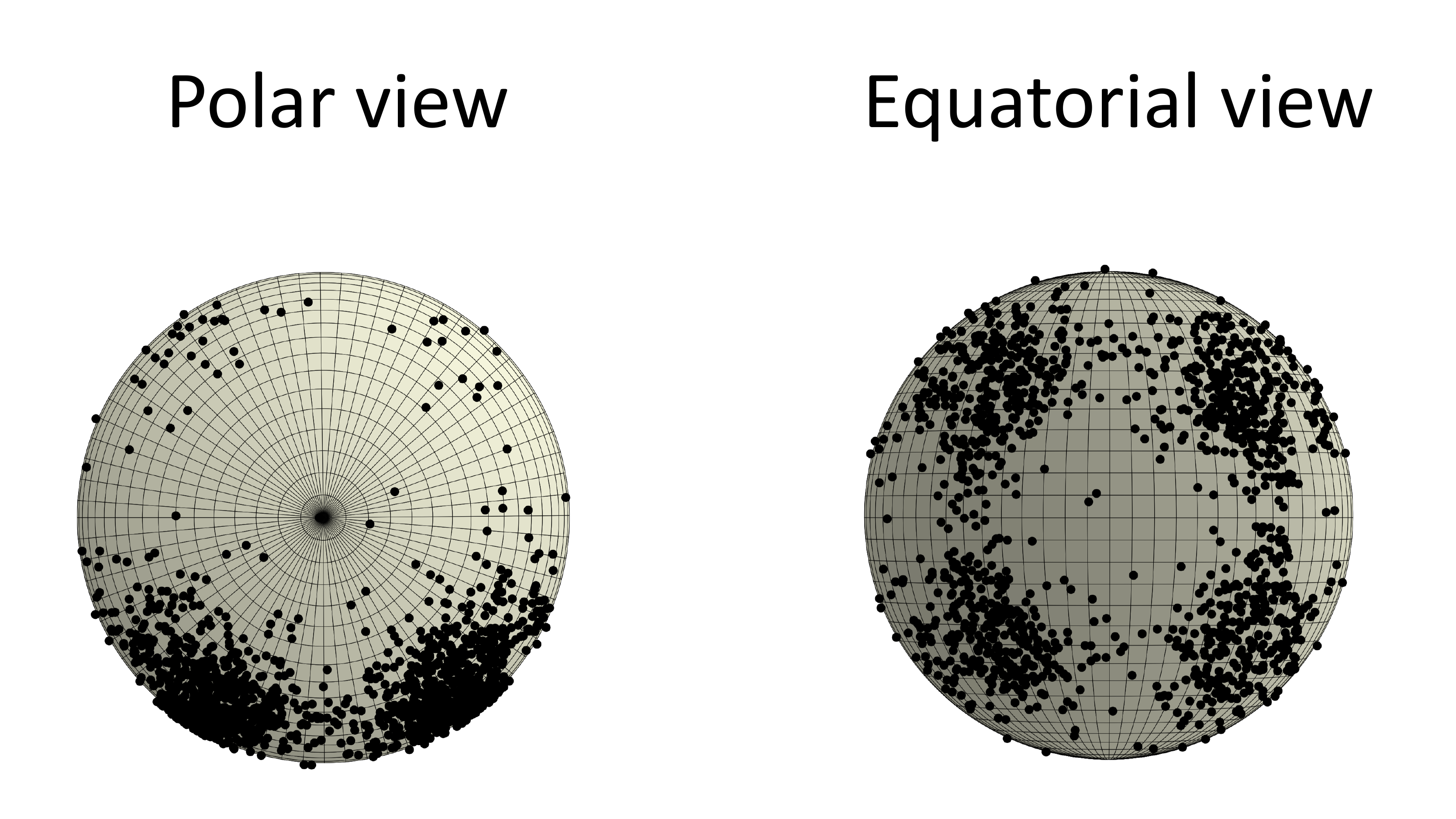}
    \caption{Similar to Figure \ref{fig:disc_surface} but for the impacts resulting from high-inclination Neptune-centric impactors viewed from the pole (left panel) and from the equator facing the leading hemisphere (right panel).}
    \label{fig:irregular_surface}
\end{figure}

This population of impactors result in an interesting distribution of impacts on the surface of Triton as seen from Figure \ref{fig:irregular_surface}.
When Triton is viewed from the front, we see that impacts are concentrated at four ``islands'' on the leading hemisphere.
These ``islands'' are distributed symmetrically about the equator and the meridian.
There are relatively fewer impacts in the equatorial and meridional region, as well as the region surrounding the apex.
The polar view shows that this population impactors, similar to the prograde, low-inclination Neptune-centric impactors, also strike Triton preferentially on its leading hemisphere.
The peculiar impact distribution is due to the high eccentricity and inclination of the orbits of this impactor population.
A higher orbital eccentricity results in a more even distribution of craters and a higher orbital inclination results in craters occurring at higher latitudes away from the equator.

\begin{figure}
	\includegraphics[width=\columnwidth]{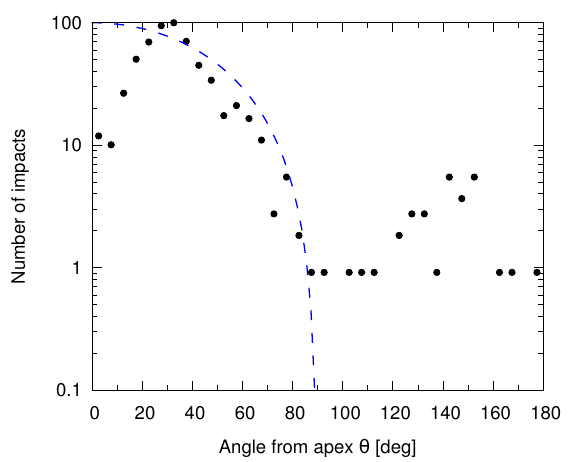}
    \caption{Projection of the impact distribution produced by high-inclination Neptune-centric impactors onto the $xy$-plane.}
    \label{fig:irregular_xy}
\end{figure}

Figure \ref{fig:irregular_xy} shows a comparison of the shape of the impact distribution on the $xy$-plane with the results of SZ07.
The peak of the impact distribution produced by high-inclination Neptune-centric impactors is displaced about 30$^{\circ}$ from the apex, in contrast to the observed crater distribution where the peak is located at the apex. 
The shift of the peak away from the apex can be computed analytically using \"{O}pik's formalism \citep{opik1976}.
For simplicity we assume that all collisions with Triton are head-on.
In that case the velocity and radius vector of the impactor with respect to Triton are anti-parallel and we can use the direction of the velocity vector to pinpoint the location of impact on Triton's surface.
\cite{carusietal1990} gave the relation between the components of the relative velocity vector between the impactor and Triton ($U_x, U_y, U_z$) to the orbital elements of the impactor.
\begin{equation}
  \begin{aligned}
      U_x &= \left[ 2-\frac{a_T}{a}-a(1-e^2) \right] ^{1/2}, \\
          &\approx \sqrt{2\left( 1- \frac{q}{a_T} \right)}, a \gg a_T, \text{and}\ q \ll a,
  \end{aligned}    
\end{equation}

\begin{equation}
  \begin{aligned}
      U_y &= \sqrt{a(1-e^2)}\cos I -1, \\
          &\approx \sqrt{2\frac{q}{a_T}}\cos I -1,
  \end{aligned}    
\end{equation}

\begin{equation}
  \begin{aligned}
      U_z &= \sqrt{a(1-e^2)}\sin I, \\
          &\approx \sqrt{2\frac{q}{a_T}}\sin I,
  \end{aligned}    
\end{equation}
where the orbital elements without a subscript are those of the impactor.
We simplify the equations by assuming $q/a_T = 1/2$, so that $U_x = 1$, $U_y = \cos I -1$, and $U_z = \sin I$. 
Taking $I = 110^{\circ}$ as an example, the angles of the impactor velocity vector measured from the apex ($+y$-axis, where $\theta_v = 0^{\circ}$) when projected onto the $xy$- and $yz$-plane are
\begin{equation}
    \begin{aligned}
       \theta_{v,xy} &= \arctan \left( \frac{U_x}{U_y} \right),\\
       &= -37^{\circ},
    \end{aligned}
\end{equation}
and
\begin{equation}
    \begin{aligned}
       \theta_{v,yz} &= \arctan \left( \frac{U_z}{U_y} \right),\\
       &= -35^{\circ},
    \end{aligned}
\end{equation}
respectively.
An impactor with $I = 110^{\circ}$ will strike the surface of Triton at $37^{\circ}$ away from the apex and $35^{\circ}$ away from the equator.
Two islands are created from impacts on the incoming leg of the impactor's orbit; the other two occurred during the outgoing leg.
Together with the two nodes of the orbit this results in four impact islands.
Although Neptune-centric impactors with high orbital inclination results in an interesting impact distribution, they too could not account for the observed crater distribution on Triton.

\subsubsection{Heliocentric impactors with hyperbolic orbits}
\begin{figure}
	\includegraphics[width=\columnwidth]{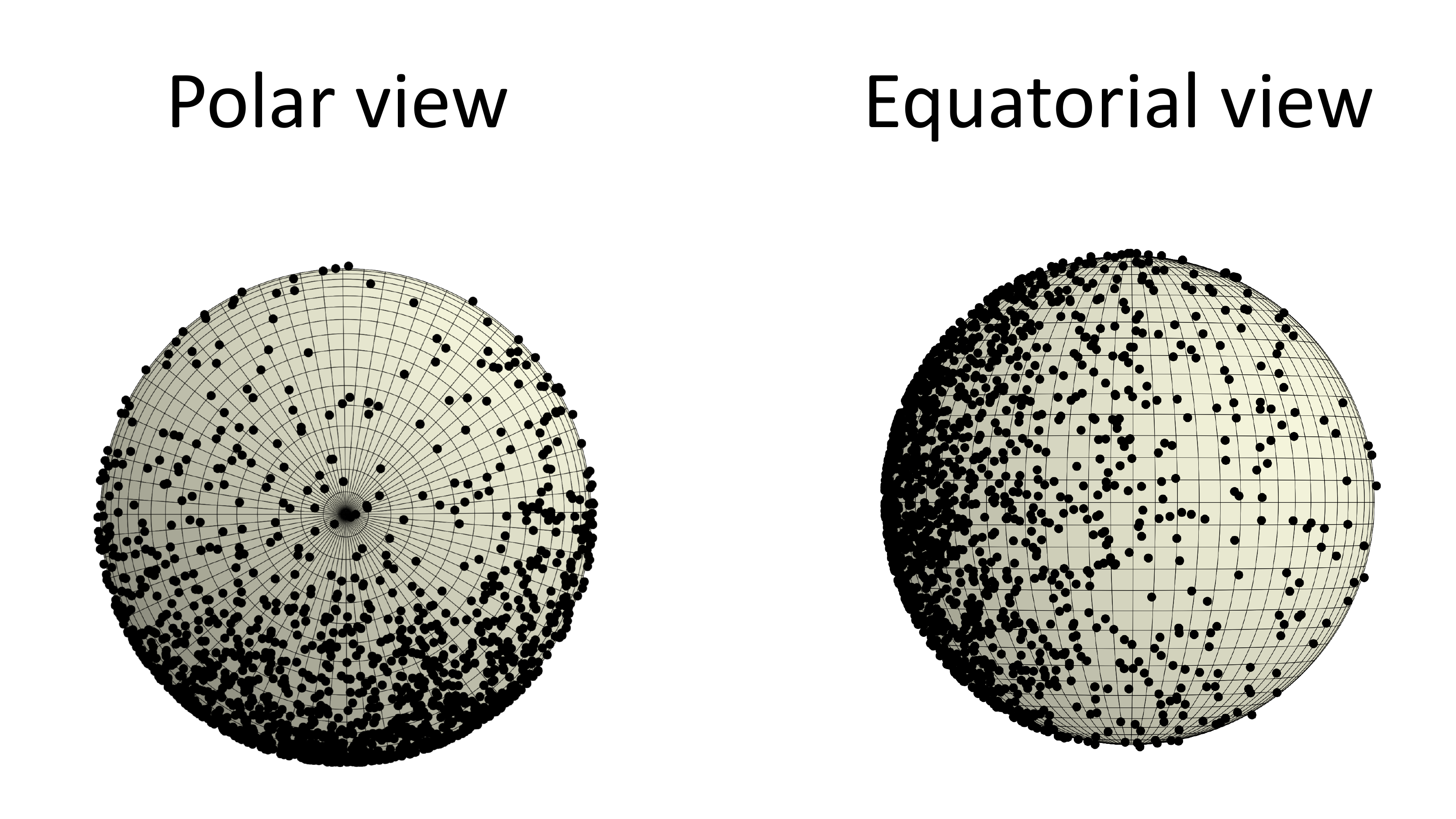}
    \caption{Similar to Figure \ref{fig:disc_surface} but for impacts produced by heliocentric impactors with hyperbolic orbits viewed from the pole (left panel) and from the equator where the leading hemisphere is on the left (right panel).}
    \label{fig:hyperbolic_surface}
\end{figure}

This is a population of impactors with hyperbolic orbits of large semi-major axes.
The impactors also have a wide range of values for their orbital inclinations because they are isotropic with respect to Neptune.
As such, they will impact Triton from all directions.
The outcome of the numerical simulations supports this.
Figure \ref{fig:hyperbolic_surface} shows that impacts by a hyperbolic population occur on the whole surface of Triton.
There is an asymmetry in the distribution with more impacts preferentially occurring on the leading hemisphere due to Triton being synchronously locked to Neptune.
However, the degree of the asymmetry in this case is less extreme compared to that due to the prograde, low-inclination Neptune-centric impactor population.

\begin{figure}
	\includegraphics[width=\columnwidth]{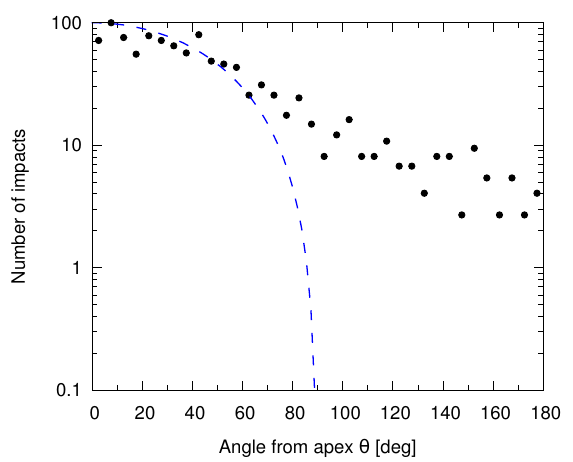}
    \caption{Projection of the impact distribution produced by heliocentric impactors with hyperbolic orbits onto the $xy$-plane.}
    \label{fig:hyperbolic_xy}
\end{figure}

As with the case for the Neptune-centric impactors (both low- and high-inclination), the hyperbolic impactors similarly could not reproduce the impact distribution as reported in SZ07.
In contrast to the steep cutoff at $\theta = 90^{\circ}$ for the observed crater distribution, the results from our numerical investigations show that hyperbolic impactors produce a very even impact distribution with the number of impacts decreasing gradually with distance from the apex (Figure \ref{fig:hyperbolic_xy}).

The results from our \textit{N}-body simulations indicate that neither of the three impactor populations, if taken independently, could reproduce the observed asymmetry of crater distribution on Triton.
The impact distribution produced by all three impactor populations are inconsistent with Triton's surface crater distribution.
The degree of discrepancy varies with each population but all the impactors strike Triton on its trailing hemisphere to various extents.
This implies that Triton's craters could not be the outcome of impacts from a single impactor population.

\subsection{Heterogeneous impactor populations}
\begin{figure*}
	\includegraphics[width=\textwidth]{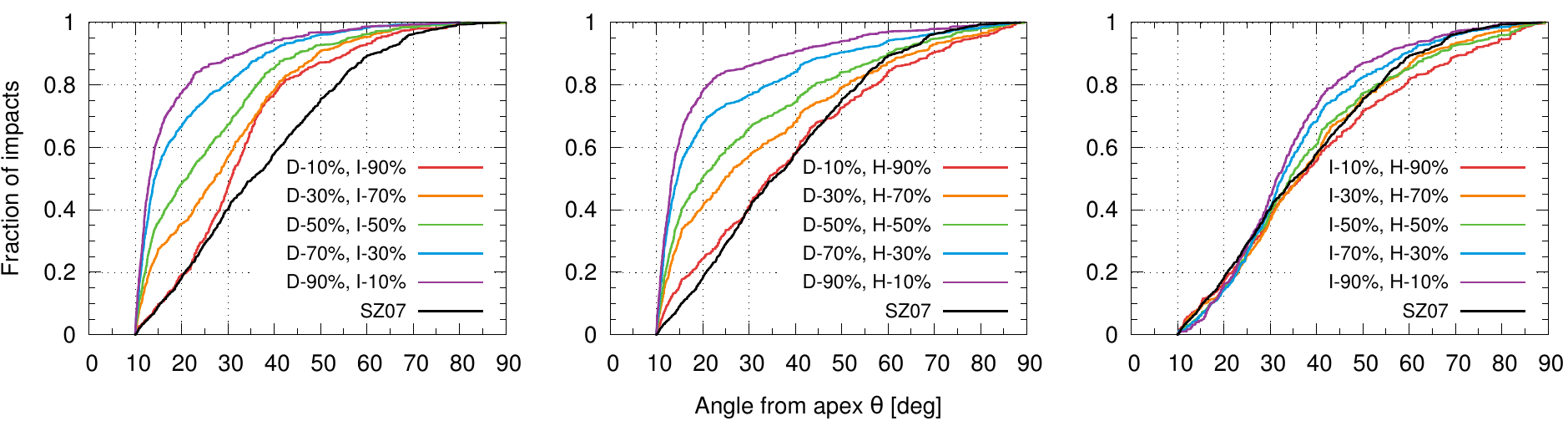}
    \caption{Cumulative distribution of the fraction of impacts with angle from the apex for mixed impactor populations consisting of prograde, low-inclination Neptune-centric (D) and high-inclination Neptune-centric (I) impactors (left panel), low-inclination Neptune-centric and heliocentric (H) impactors (middle panel), and high-inclination Neptune-centric and heliocentric impactors (right panel). Coloured lines depict the various mixing ratios of each population. Black line is the distribution of impacts obtained from the observed crater density distribution reported in SZ07.}
    \label{fig:convolution_2}
\end{figure*}

\begin{figure}
	\includegraphics[width=\columnwidth]{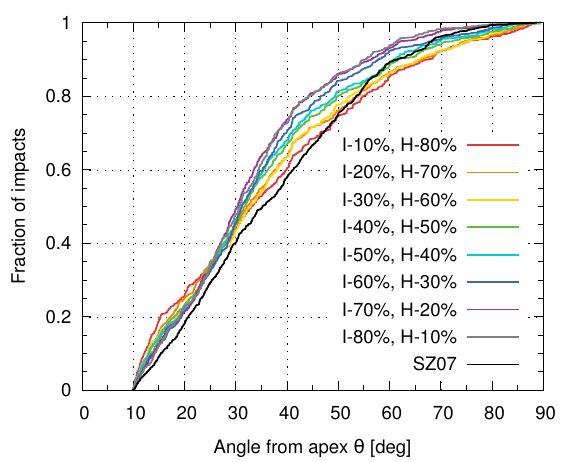}
    \caption{Same as Figure \ref{fig:convolution_2} but for a mixed impactor population consisting of 10\% prograde, low-inclination Neptune-centric impactors and various mixing ratios of high-inclination Neptune-centric and heliocentric impactors.}
    \label{fig:convolution_3}
\end{figure}

Were Triton's craters produced by a mixed population of impactors instead?
We tested this hypothesis by comparing impact distributions resulting from a combination of two impactor populations with different mixing ratios, generated using a Monte Carlo approach, with the impact distribution obtained from the observed crater density reported in SZ07.
The outcome is presented in Figure \ref{fig:convolution_2}.
We plot the cumulative fraction of impacts against impact angle for various mixing ratios of different impactor populations.
Qualitatively, the combination of impacts from a mixed impactor source comprising of high-inclination Neptune-centric impactors and heliocentric impactors result in an overall impact distribution that matches the observations best.
We further subjected these impact distributions to a Kolmogorov-Smirnov (K-S) test to determine their goodness of fit.
The results of the K-S test indicates that the impact distribution produced by a mixed population of high-inclination Neptune-centric (I) impactors and heliocentric (H) impactors with mixing ratio I:H = 30\%:70\% fits the observed crater density distribution best with a K-S statistic of 0.051.
The probability that both distributions (the Monte Carlo best-fit and the observed crater density) match is 14\%.

Given the combination of high-inclination Neptune-centric and heliocentric impactors with best-fit mixing ratio of 30\%:70\%, we went a step further and included the low-inclination Neptune-centric impactors into the mix to see if it would yield a better fit with a lower K-S statistic.
In Figure \ref{fig:convolution_3} we show the resulting impact distributions when the low-inclination Neptune-centric impactors account for 10\% of the mixed impactor population.
The best-fit according to the results of the K-S test is D:I:H = 10\%:30\%:60\% with a K-S statistic of 0.074.
It appears that including low-inclination Neptune-centric impactors into the mix does not aid in improving the fit to the observed distribution.
Considering that there is no viable mechanism to generate such a population of prograde, low-inclination impactors such a short time ago, we can rule out the contribution of low-inclination Neptune-centric impactors to the craters on Triton.
Our results suggest that the craters on Triton are products of impacts due to the combination of a high-inclination Neptune-centric source with orbits akin to the Neptunian irregular satellites, and a heliocentric source.

We compute an estimate of the surface age of Triton from our results.
\cite{levisonetal2000} computed the impact probability of the scattered disc objects (SDOs) with the giant planets and found that the impact probability with Jupiter is 0.57\%.
The relative impact probability with Triton when scaled to Jupiter is 7.5$\times10^{-5}$ \citep{zahnleetal2003}.
The product of the two probabilities gives the impact probability of the SDOs with Triton as 4.3$\times10^{-7}$.
An independent estimate by \cite{bm2013} gave the leakage rate of comets from the SD as 1.66$\times10^{-10}$/yr and the number of SDOs with diameter $D > 2.3$ km as 2$\times10^9$, resulting in 0.33 comets leaking from the SD per year.
Combining this with the impact probability, we estimate that Triton receives an impact from a comet larger than 2.3 km every 7 Myr.
A comet of $D$ = 2.3 km creates a crater with a diameter of about 28 km.
As there are about 12 such craters on Triton (SZ07), the surface age is about 84 Myr if all of the impacts originate from a heliocentric source, and less than 63 Myr if 30\% of the impacts are due to a high-inclination Neptune-centric source as indicated by our best-fit result.
The surface age estimated using our best-fit result is consistent with that estimated by SZ07.

From the impact probability obtained from the results of our \textit{N}-body simulations, we could in principle compute the estimated number of Neptune's irregular satellites with $D > 2$ km. 
Given that the irregular satellites produce about 3 craters with $D > 28$ km (9 are produced by comets) and an impact probability of 6\%, there are roughly 50 irregular satellites which encountered Triton over the course of 63 Myr. 
Assuming a constant flux, we find that there are about 3500 irregular satellites with $64^{\circ} \leq I \leq 110^{\circ}$ out of a total of 16\,000.
\cite{bottkeetal2010} computed the number of irregular satellites for Uranus (which can serve as a proxy for Neptune) with $D \sim 2$ km to be $\sim10^3$.
Although our estimate is higher, it is reasonable considering the uncertainties and our simple assumption of a constant impactor flux.

SZ07 remapped the craters on Triton using \textit{Voyager} images of higher quality.
Although images of only 35\% of the surface of Triton were available, the authors argued that the extreme cratering asymmetry between the leading and trailing hemisphere of Triton is real based on the absence of craters on (i) the boundary of the leading and trailing hemispheres where the resolution is the highest, and (ii) the trailing hemisphere where the resolution of the images permits identification of craters should they be present.
They conclude that impacts due to a prograde, Neptune-centric population accounts for the crater distribution on Triton more satisfactorily.
The results of our dynamical and Monte Carlo study are in disagreement with this conclusion.
When we take into account the impact dynamics of a prograde, low-inclination Neptune-centric population with Triton, the resulting impact distribution is much narrower with most of the impacts occurring within 20$^{\circ}$ from the apex.
In addition, the results of a Monte Carlo approach show that the contribution of this population of impactors to the crater distribution on Triton's surface is low, if at all.

\section{Conclusions}
\label{sec:conclusions}
Triton's extreme cratering asymmetry presents itself as an intriguing puzzle.
We adopted a dynamical approach to this puzzle by simulating the dynamics of impacts due to a prograde, low-inclination Neptune-centric population, a high-inclination Neptune-centric population, and a heliocentric population with hyperbolic orbits on Triton using a \textit{N}-body simulations.
We found that the distribution of impacts produced by each of these populations separately could not reconcile with the observed crater distribution on Triton as reported in SZ07.
A prograde, low-inclination Neptune-centric impactor population produces impacts that are concentrated within 20$^{\circ}$ from the apex, much narrow compared to the observations.
A high-inclination Neptune-centric impactor population tends to impact Triton away from the apex, resulting in an impact distribution that is qualitatively different from the observations.
A heliocentric impactor population results in a crater distribution that exhibits a leading-trailing asymmetry less extreme compared to that derived from the observations.
Based on the results of our numerical investigations, we conclude that Triton's craters could not be due solely to a single source of impactors.

The impact distribution produced by a mixed population of high-inclination Neptune-centric impactors and heliocentric impactors with a mixing ratio of 30\%:70\% appear to fit the observed crater density distribution better within the observed range of apex angle $10^{\circ} \leq \theta \leq 90^{\circ}$.
The total absence of craters on Triton's trailing hemisphere however, remains unexplained as our results show that the impactors, regardless of their source, impact Triton from the back to various extents.

\section*{Acknowledgements}
The authors are grateful for the constructive comments and reviews by an anonymous reviewer.
The simulations were run on the computing cluster of the Academia Sinica Institute of Astronomy and Astrophysics in Taiwan when JM was a participant of their Summer Student Programme.




\bibliographystyle{mnras}
\bibliography{triton_crater_asymmetry} 


\bsp	
\label{lastpage}
\end{document}